\documentclass[journal]{IEEEtran}
\usepackage{graphicx}
\usepackage{wrapfig}
\usepackage{gensymb}
\usepackage[utf8]{inputenc}
\usepackage{jsen}
\usepackage{scalerel}
\usepackage{tikz}
\usetikzlibrary{svg.path}

\definecolor{orcidlogocol}{HTML}{A6CE39}
\tikzset{
	orcidlogo/.pic={
		\fill[orcidlogocol] svg{M256,128c0,70.7-57.3,128-128,128C57.3,256,0,198.7,0,128C0,57.3,57.3,0,128,0C198.7,0,256,57.3,256,128z};
		\fill[white] svg{M86.3,186.2H70.9V79.1h15.4v48.4V186.2z}
		svg{M108.9,79.1h41.6c39.6,0,57,28.3,57,53.6c0,27.5-21.5,53.6-56.8,53.6h-41.8V79.1z M124.3,172.4h24.5c34.9,0,42.9-26.5,42.9-39.7c0-21.5-13.7-39.7-43.7-39.7h-23.7V172.4z}
		svg{M88.7,56.8c0,5.5-4.5,10.1-10.1,10.1c-5.6,0-10.1-4.6-10.1-10.1c0-5.6,4.5-10.1,10.1-10.1C84.2,46.7,88.7,51.3,88.7,56.8z};
	}
}

\newcommand\orcidicon[1]{\href{https://orcid.org/#1}{\mbox{\scalerel*{
				\begin{tikzpicture}[yscale=-1,transform shape]
				\pic{orcidlogo};
				\end{tikzpicture}
			}{|}}}}
\usepackage{hyperref} 
\usepackage{epsfig}
\usepackage{epstopdf}
\usepackage{amsmath}
\usepackage{amssymb}
\usepackage{bm}
\usepackage{graphicx} 
\usepackage{booktabs}

\usepackage[font=footnotesize,skip=4pt]{subcaption}
\usepackage[font=footnotesize,skip=2pt]{caption}
		
\usepackage{jsen}
\usepackage{cite}
\usepackage{amsmath,amssymb,amsfonts}
\usepackage{algorithmic}
\usepackage{graphicx}
\usepackage{textcomp}
\usepackage{wrapfig}
\usepackage{multirow,bigstrut}

\usepackage{algorithm}
\usepackage{algorithmic}
\usepackage{lipsum}
\usepackage{graphicx}
\usepackage{siunitx}
\usepackage{soul}

\def\BibTeX{{\rm B\kern-.05em{\sc i\kern-.025em b}\kern-.08em
		T\kern-.1667em\lower.7ex\hbox{E}\kern-.125emX}}

\begin{document}

\author{Maxim Popov, Temirgali Aimyshev, Eldar Ismailov, Ablay Bulegenov, Siamac Fazli
\thanks{M. Popov and S. Fazli are with the Department of Computer Science, Nazarbayev University, Astana, Kazakhstan. T. Aimyshev, E. Ismailov and A. Bulegenov are with CMC Technologies, Astana, Kazakhstan. Corresponding author: S. Fazli (siamac.fazli@nu.edu.kz). The co-authors would like to acknowledge the support of Nazarbayev University Research Grants funding (240919FD3926) and .
}}

\title{A Review of Modern Approaches for Coronary Angiography Imaging Analysis}

\maketitle

\begin{abstract}
Coronary Heart Disease (CHD) is a leading cause of death in the modern world. The development of modern analytical tools for diagnostics and treatment of CHD is receiving substantial attention from the scientific community. Deep learning-based algorithms, such as segmentation networks and detectors, play an important role in assisting medical professionals by providing timely analysis of a patient's angiograms. This paper focuses on X-Ray Coronary Angiography (XCA), which is considered to be a ``gold standard'' in the diagnosis and treatment of CHD. First, we describe publicly available datasets of XCA images. Then, classical and modern techniques of image preprocessing are reviewed. In addition, common frame selection techniques are discussed, which are an important factor of input quality and thus model performance. In the following two chapters we discuss modern vessel segmentation and stenosis detection networks and, 
finally, open problems and current limitations of the current state-of-the-art.
\end{abstract}

\begin{IEEEkeywords}
Computer Vision, Artificial Intelligence, Image Segmentation, Image Classification, Object Detection, Coronary Artery Disease, Coronary Heart Disease, X-ray Coronary Angiography, Image Preprocessing, Deep Learning, Object Tracking, Review.
\end{IEEEkeywords}

\section{Introduction}
Coronary Heart Disease (CHD), also known and referred to as Coronary Artery Disease (CAD), is among the most common causes of death in the world~\cite{mccullough2007coronary}. It is characterized by the presence of atherosclerotic plaques in coronary vessels~\cite{ralapanawa2021epidemiology}. According to different sources, coronary artery disease takes up to 19 million lives annually, constituting for about $32\%$ of global mortality~\cite{ferreira2014epidemiology}. In particular, according to the survey on heart disease and stroke in the US, about 15.4 million people older than 20 suffer from ischemic heart disease~\cite{go2013ihdsurvey}. 

X-Ray Coronary Angiography (XCA) is a gold standard in the diagnosis and treatment of CAD~\cite{barbanti2017optimized, Carballal2018AutomaticMV}. In this method, contrast agents (e.g. iodixanol-320) are injected to coronary vessels through a catheter to highlight the blood flow on x-ray angiograms. It enables clinicians to obtain detailed images of coronary arteries. These angiograms are used by medical professionals to make qualitative judgements about the vascular wall calcification and stenosis problems~\cite{LIANG2021102894, ZHU2021105897}.

Manual and semi-automatic analysis of angiograms requires significant amount of time, thus reducing the number of patients receiving treatment. In addition it relies heavily on the doctor's personal experience and capability, thus resulting in the issue of inter- and intra-observer variability~\cite{Carballal2018AutomaticMV}. Inter-observer variability means that annotations done by several different doctors may have diverging labels and region borders. Intra-observer variability means that the same doctor may, on different occasions, produce inconsistent annotations. Reasonable attempts to automate this process in order to overcome above mentioned limitations using different algorithms have been made over the last 20 years~\cite{schaap2011robust, brieva2006level, bock2008robust, khan2006facilitating, li2012novel, wahle2006plaque, zhou2008new, kose2006fully}, and especially blossomed with the emergence of the first fully convolutional neural networks~\cite{krizhevsky2012} and their further development in the form of segmentation networks~\cite{deeplabv3plus2018,liu2019autodeeplab, wang2020maxdeeplab, qiao2021vipdeeplab, cheng2020panopticdeeplab, wang2020axialdeeplab, yu2022cmtdeeplab, meng2022unet3plus, Huang2020unet3plus,zhou2018unet++,zhang2018resunet,cciccek20163dunet,isensee2018nnunet,oktay2018attentionunet,ronneberger2015unet, chen2017deeplab}. Convolutional neural networks (CNNs) have achieved remarkable results in many medical applications, including the detection of Covid-19~\cite{ozturk2020automated}, cancer~\cite{grovik2021deeplabbrain,negi2020unetbrest}, retinopathy~\cite{li2019mresunetretinal}, and many other serious diseases, given the public availability of data and pretrained models.

However, to date, a number of unsolved problems prevail, which hinder accurate and efficient diagnosis of medical X-ray coronary angiograms by means of automated algorithms. These problems include borders from X-ray filters, rotation of the image frames, varying levels of contrast agent, magnification during image acquisition~\cite{Iyer2021angionet}, noisy image capture, sample/patient variability~\cite{SAZAK2019739} and deformation of the shape of blood vessels on the image~\cite{LIANG2021102894}. While a number of solutions have recently been proposed, which alleviate some of these issues, others are still unsolved at this time. Unfortunately, however, a comprehensive review of open issues and their potential resolution is currently lacking within the scientific literature and this review attempts to fill this gap.

\section{Open Datasets}
One of the most important issues in this area is the unavailability of large and reliable datasets for model estimation and testing, similiar to those which exist for retinal images – DRIVE~\cite{drive2004} and STARE~\cite{hoover2000stare}. Currently, only very few datasets are publicly available for benchmarking. For example, \cite{sanchez2019ann} provides 134 angiograms and their respective ground truth images labelled by professional clinicians for open access. Another paper also included 30 4-frame sequences from angiographic videos and corresponding binary masks~\cite{hao2020svsnet}. One challenge facing the production of ready-to-use datasets for this task lies in the fact that not all images and XCA sequences are suitable for segmentation. Vessel visibility due to the propagation of contrast agent, as well as the cardiac phase are playing an important role for medical specialists in determining a suitable frame for analysis. Another challenge is given by the low availability of certified medical professionals, including cardiologists and radiologists, for reliable dataset production.

\section{Image Preprocessing Methods}
Image preprocessing is crucial for the automated analysis of angiographic images. As it was stated before, the domain of angiography is intertwined with highly noisy images, patient’s motions, varying levels of contrast, thus vessel segmentation and lesion detection tasks require some level of image enhancement, so that visual characteristics of the image can be upgraded. Image enhancement techniques can be split into 2 parts, with the first one being morphological operations such as dilation and erosion~\cite{Mulay2021adain, gao2022vessel,SAZAK2019739}, and second one being advanced mathematical algorithms like Multiscale Retinex with Color Restoration (MSRCR)~\cite{Rahman1998msrcr} and Contrast Limited Histogram Equalization (CLAHE)~\cite{pisano1998contrast} and even specialized deep neural networks~\cite{Iyer2021angionet}.

\begin{figure*}[ht]
	\begin{center}
	\includegraphics[width=\textwidth]{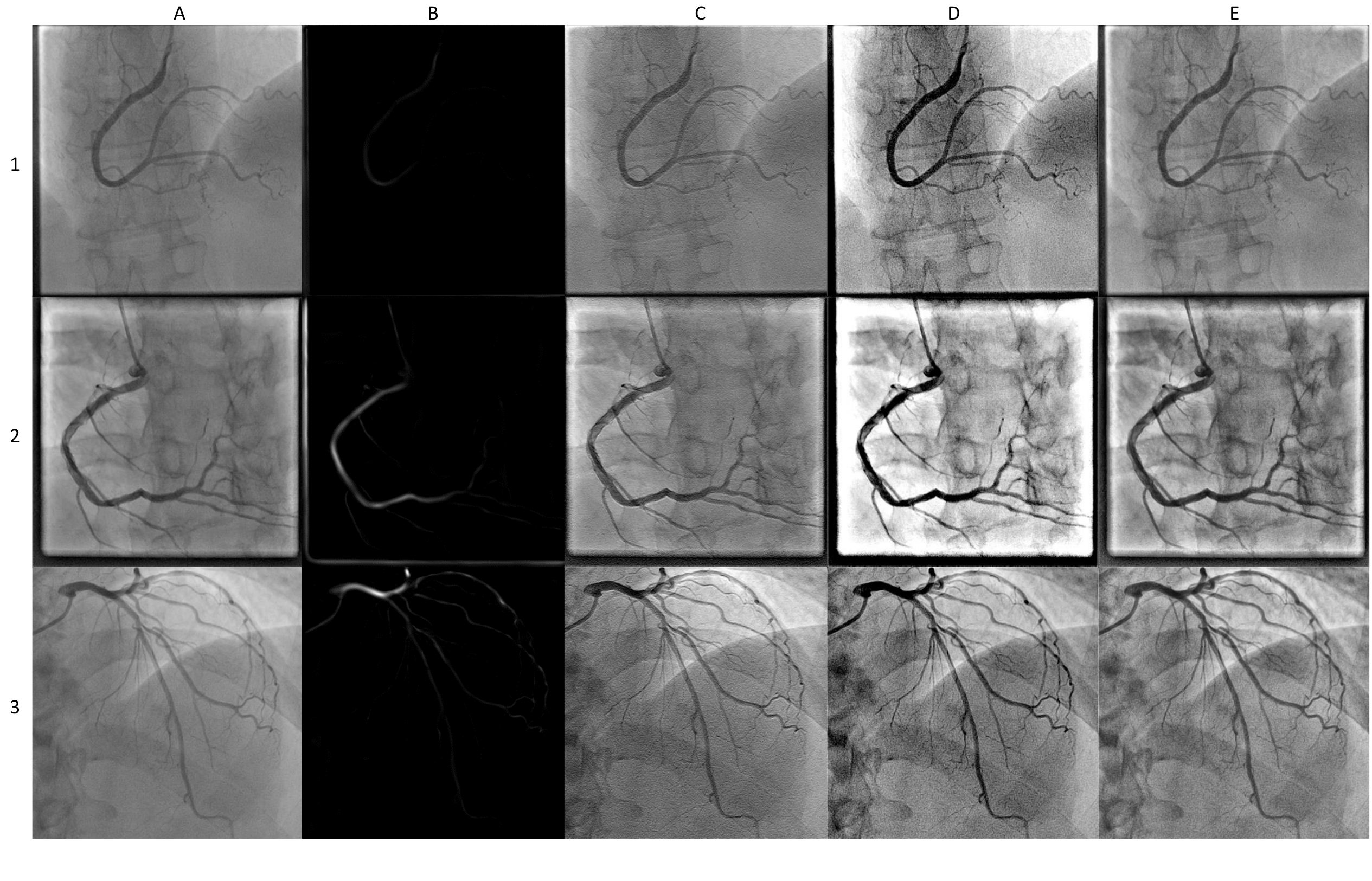}
	\caption{Gives an overview of various preprocessing methods in cornary angiography. A1-A3. Shows raw images; B1-B3. Shows the result of application of Hessian-based Frangi Vesselness filter; C1-C3. Shows images after the top-bottom hat transform was applied; D1-D3. Shows the result of applying MSRCR; E1-E3. Shows the results of CLAHE on test images.}
	\label{fig:prepro}
	\end{center}
\end{figure*}

The first example is the top-bottom hat transform, as can be seen in the third column of Fig.~\ref{fig:prepro}C. The top-bottom hat transform is one of the most common and computationally simple tools for edge visibility enhancement. It is based on multiscale high- and low-pass filters. High-pass filters leave only the bright spots, which usually represent vessels on the image, while low-pass filters aim to leave only the background of the image. The resulting filtered image is obtained by the following formula \cite{Mulay2021adain}: 
\begin{equation}
I_{en} = I + I_{th} - I_{bh},
\end{equation}
where $I_{en}$ is the enhanced image, $I$ the original image, $I_{th}$ the image after the top-hat transform, and $I_{bh}$ the image after the bottom-hat transform. A slightly altered version of this transform includes adjustable weight parameters for $I_{th}$ and $I_{bh}$~\cite{gao2022vessel}.

Contrast Limited Adaptive Histogram Equalization (CLAHE) is an image preprocessing algorithm~\cite{pisano1998contrast}, for improved visibility of lesions on mammogram images. This method is based on the application of Adaptive Histogram Equalization (AHE) on small regions of the image, called tiles, instead of the whole image. Individual tiles are then combined using interpolation techniques, in order to remove artificial boundaries, occurring after equalization. Currently, this method is widely applied in retinal image segmentation~\cite{setiawan2013claheretinal}, chest x-ray images~\cite{reamaroon2020robust}, and also found in the application of XCA segmentation~\cite{JIANG2021100602} An example of CLAHE on coronary angiograms can be found in the fifth column of Fig.~\ref{fig:prepro}E.

Multi-scale Retinex with Color Restoration (MSRCR) is an algorithm designed to remove the artifacts from the image while preserving sharpness on the edges by providing color constancy and dynamic color range suppression~\cite{Rahman1998msrcr}. It is a computationally efficient tool and results in the improvement of the image structure, effectively removing noise from the angiogram. It is often considered as an important step in the preparation of images for further preprocessing methodologies such as~\cite{frangi98hessian}, since this methodology is known to be highly sensitive to small artifacts left by noise. Further development of classic MSRCR architecture was proposed in~\cite{zhang2022bettermsrcr}. Zhang and colleagues had utilized MSRCR in a pipeline, consisting of Singular Value Decomposition in the RGB color space, automated MSRCR and Gamma Stretching for luminance and saturation enhancement in the HSV color space, and Guided Filtering for final denoising in RGB space. Examples of application of classical MSRCR on XCA images can be found in~\cite{Liu2016msrcr} and on Fig.~\ref{fig:prepro}D.

Hessian filters are an example of computationally expensive mathematical image enhancement techniques. This method was initially designed to detect blob-like structures from local maxima of a given image~\cite{frangi98hessian}. A Hessian filter for a 2D image is a 2x2 matrix, composed of second order partial derivatives of the intensity function of the image.

\begin{equation}
    H_2 (x) = \begin{pmatrix}
        \frac{\delta^2f}{\delta x^2} & \frac{\delta^2f}{\delta x\delta y} \\
        \frac{\delta^2f}{\delta y\delta x} & \frac{\delta^2f}{\delta y^2}
    \end{pmatrix}
    \triangleq 
    \begin{pmatrix}
    f_{xx} & f_{xy} \\
    f_{yx} & f_{yy}
    \end{pmatrix}
\end{equation}

Reconstructed images with highlighted tubular structures $I_{\sigma}(\overline{x})$ are calculated separately for different values of $\sigma$ by convolution of the original image $I$ with a Gaussian kernel $G_\sigma(\overline{x})$:
\begin{equation}
    I_\sigma(\overline{x}) = I * G_\sigma(\overline{x})
\end{equation}
The value for $\sigma$ is selected from range $0.5$ to $6$ times the system resolution of $165\mu$m, usually being equal to $2$ to $10$ pixels. 
The normalized Hessian matrix of the resulting image $I_\sigma(\overline{x})$ for the scale $\sigma$ is given by:
\begin{equation}
    H(\overline{x},\sigma) = B \frac{\delta^2I_\sigma(\overline{x})}{\delta\overline{x}^2}
\end{equation}
With B being a normalization factor.

The eigenvectors of a Hessian matrix define the principal directions of local image features. Blood vessel regions can be detected based on the properties of eigenvalues ($\lambda_1$,$\lambda_2$), with a special condition $|\lambda_1| \leq |\lambda_2|$. The Hessian-based Frangi Vesselness filter is used to extract tubular structures from the image using the vesselness function~\cite{frangi98hessian}. Frangi~\emph{et al.} specify the vascular similarity formula for a given set of eigenvalues as follows:
\begin{equation}
V_f(\overset{\rightarrow}{\lambda}) = \Biggl\{
\begin{aligned}
& ~ 0, \qquad \text{if} \quad \lambda_2 < 0 \\
&\exp{\left(-\frac{R^2_{\beta}}{2\beta^2}\right)} \left( 1-\exp{(1-\frac{-S^2}{2\gamma^2}}) \right)
\end{aligned}
\end{equation}

$R_\beta$ here is the blobness measure, used to distinguish vessels from blob-like structures, and calculated by $\frac{\lambda_1}{\lambda_2}$. $S$ is the second order structureness measure, used to distinguish vessels from the background and noise, calculated by $S = \sqrt{\lambda_1^2+\lambda_2^2}$. Parameters $\beta$ and $\gamma$ are weights that measure the influence of $R_B$ and $S$ on the vascular similarity response. This formula is used to reconstruct a final vessel-enhanced image at different scales $\sigma$ by summing filter responses. This way, vesselness function returns the value in range $(0, 1)$, with maximum value corresponding to a tubular structure of scale $\sigma$, and smallest value corresponding to background pixels~\cite{frangi98hessian}.

While having been discovered several decades ago, Hessian filters are still commonly employed with new modifications being released up to this date~\cite{zhao2019ratvasculature}. In fact, they remain one of the most useful and popular baseline methods for unsupervised image enhancement and segmentation to this day~\cite{tsai2015automatic, zai2018effective, ansari2016performance,cruz2015automatic}. Practical applications of Hessian filters for tubular structures detection can be observed on the Fig.~\ref{fig:prepro}B.

Sazak and colleagues in their work provide criticisms for popular enhancement methods, such as Hessian-based, mathematical morphology-based, and adaptive histogram-based filters. As an alternative, they propose an improvement over the method proposed earlier by Zana and Klein~\cite{zana2001morph}, called multiscale bowler-hat transform~\cite{SAZAK2019739}. This approach is based on mathematical morphology and uses 2 morphological operations to perform vessel enhancement: erosion and dilation. Dilation ($\oplus$) for a given pixel $I(p)$ with a structuring element $b(p)$ in a greyscale image is defined as:
\[
(I \oplus b)(p) = \underset{x \in E}{sup}[I(x) + b(p-x)]
\]
where “sup” is the supremum and $x \in E$ denotes all points in the Euclidean space of the image. Erosion ($\ominus$) is defined in similar terms:
\[
(I \ominus b)(p) = \underset{x \in E}{inf}[I(x) + b(p-x)]
\]
where ‘inf’ is the infimum. These two operations are combined to form the opening operation, which is the erosion of the original image, followed by the dilation using the same structuring element. Sazak \emph{et al.} proposed to use a combination of lines with 12 different orientations and different radii, and disks with different radii as structuring elements for the opening operation. For each diameter of a vessel $d$ from $1$ to $d_max$, a disk shaped filter is used to produce a stack of images, on which any blob-like element of size smaller than $d$ are removed. The same goes for linearly structuring elements – for each combination of direction $\theta$ and length of structuring element $d$, a distinct image is produced using opening operation. The enhanced image is then formed from these two stacks, using the maximum difference at each pixel between the stacks for each $d$. This way, pixels in the background are removed using opening with disk-shaped structuring elements, and pixels in the foreground are preserved, i.e., vessel-like structures will have a high value due to line-based structuring elements.

The Angiographic Processing Network (APN) is a modern approach to image enhancement~\cite{Iyer2021angionet}. It is based on deep neural networks which are trained using images, that are enhanced with hand-crafted filters, as targets, with the aim of improving vessel visibility. The idea behind designing this network is that while the perfect filter is generally unknown, it can be learned by the network from labelled examples. APN is initialized to mimic a combination of standard enhancement filters, such as unsharp mask filters. These filters are constructed with an aim to improve vessel boundaries and local contrast on vessel edges. Then, the network is fine-tuned to produce a new filter that suits the task of image segmentation most.

In summary, image enhancements are an essential tool for the vessel segmentation task, and sometimes they can accomplish this task, without the need of applying more complex algorithms, such as deep neural networks.

\section{Best Frame Selection}
Frame selection is an important step for obtaining clean data, which is a prerequisite for developing a well-performing deep neural network. Input data may be compromised by cardiac movements, spread of color agents, noisy image capture and movements of patient, which can negatively affect the network's performance. Another important reason for frame selection is that it reduces that number of input dimensions, which makes the DNN less likely to overfit.
Frame selection is usually performed manually~\cite{diagnostics12040778, WAN2018179, gao2022vessel} either without any additional modalities or in combination with ECG signals, which are collected synchronously with XCA~\cite{cuisdel2018ecg}. In order to automate the frame selection process and thus assist radiologists, several algorithmic approaches have been proposed. They range from complex deep neural networks with million of parameters and thousands of training samples, to simpler models relying on the developer's domain knowledge.

A CNN + LSTM framework for the selection of ideal candidate frames has recently been introduced~\cite{cong2019frame}. Best frame was described as the one having the best image quality, full contrast-agent penetration, clearly contrasted vessel borders, and anatomical significance of stenosis, if present. Initially, inception-v3~\cite{szegedy2015inceptionv3} was pretrained on 1073 images from 19 patients for recognizing ideal candidates, and then its fully connected layer was connected to a pair of bi-directional LSTM networks based on~\cite{Ma_2017}. The resulting network was trained with 18688 frames from 19 patients, with videos normalized to contain 128 frames. Performance of this step was assessed on 582 XCA videos from 175 patients. 
Initially, medical professionals indicated the Beginning and Ending Constrasting Frames on the ICA sequences. These denote the time interval during which vessel colouring is reached. Then, neural network was trained to detect these key frames automatically. The result of the inference was considered as acceptable, as long as the average absolute difference between predicted and real key frames was lower than $3$.
This way, an acceptance and error rate of $0.83$ and $0.0498$ were obtained, respectively. This method is suitable for data filtering for medical professionals, however, it does not indicate which frames are suitable for the detection task, since it does not provide cardiac phase information.

U-net~\cite{ronneberger2015unet} inspired neural network that extracts temporal features from angiographic video was proposed in~\cite{cuisdel2018ecg}. This paper focuses on cardiac phase decoding based on angiographic images in order to determine the best frame for stenosis detection. This is accomplished by selecting images where the negative effects of heart movements are minimized.
For this purpose, authors designed two separate neural networks – the first one detects the time window of optimal vessel coloring, and the second one selects the optimal images, based on specific time points within the cardiac cycle. For training the network, cardiac cycles are estimated from ECG. 
Training and validation sets were composed of total of 17800 angiographic images from 3900 patients, with 80-20\% split. This method has a precision and recall of 97.4\% and 96.9\%, respectively, for the task of detecting the end-diastolic phase. 

The same task of detecting frames containing the end-diastolic phase was accomplished using the key point motion detection~\cite{meng2021frame}. The accuracy of this method reached 92.99\%, utilizing data of 31 patients (as compared to 3900 in~\cite{cuisdel2018ecg}). These results show that simple models with fewer number of parameters can achieve competitive results with relatively few samples (i.e. training data). 

To select a sequence, consisting of several consecutive images for stenosis detection, a U-net~\cite{ronneberger2015unet} based network was designed and trained with 90 XCA sequences. Individual frames from these sequences were divided into 3 classes -  zero contrast filling, partial, and full contrast filling~\cite{WU2020103657}. To compensate for the lack of training samples, custom data augmentation methods were proposed. 
In their work Wu \emph{et al.} introduced a hyperparameter N. This hyperparameter indicates how many preceding and following frames should be selected with respect to the most contrast-filled frame (resulting in $2N+1$ frames in total). 
Best candidate frame in this approach was chosen using a predicted probability of a frame to represent a class of fully-filled frames. The accuracy of key frame detection using the U-net-related architecture was reported to be $0.867$, outperforming other considered architectures, such as Multiscale Region Growing~\cite{KERKENI201649}, Frangi thresholding algorithm~\cite{frangi98hessian}, Coye’s method~\cite{coye2015vessel}, and Otsu’s thresholding method~\cite{otsu1979thresholding}. The reasoning to extract several frames instead of just one was to reduce the number of false-positive responses of the stenosis detection module by comparing the stenosis detection results across several frames.

Overall, key frame detection is a task that can be successfully completed using machine learning approaches with virtually any amount of data available. Even simple models can help automating this task, and by collecting more data or by effectively utilizing domain knowledge scientist can acquire even more sophisticated results. Automated recognition of the most suitable time window and detection of an end-diastolic cardiac phase should greatly assist cardiologists and deep learning algorithm designers in diagnostics of CAD.

\section{Segmentation}
Current state of vessel annotation requires doctors to manually segment vessel regions and classify lesions and stenosis in them, relying on methods such as the Syntax Score~\cite{neumann2018syntax}. Syntax Score is an objective method of vessel analysis and stenosis detection, which is commonly used in medical practice. In recent years, the use of deep learning based neural networks for the automation of biomedical imaging diagnostics has become a viable alternative. However, these methodologies have not yet witnessed large-scale adoption and thus remain in a proof-of-concept stage.

\begin{figure}[ht]
    \centering
    \includegraphics[width=\textwidth/2]{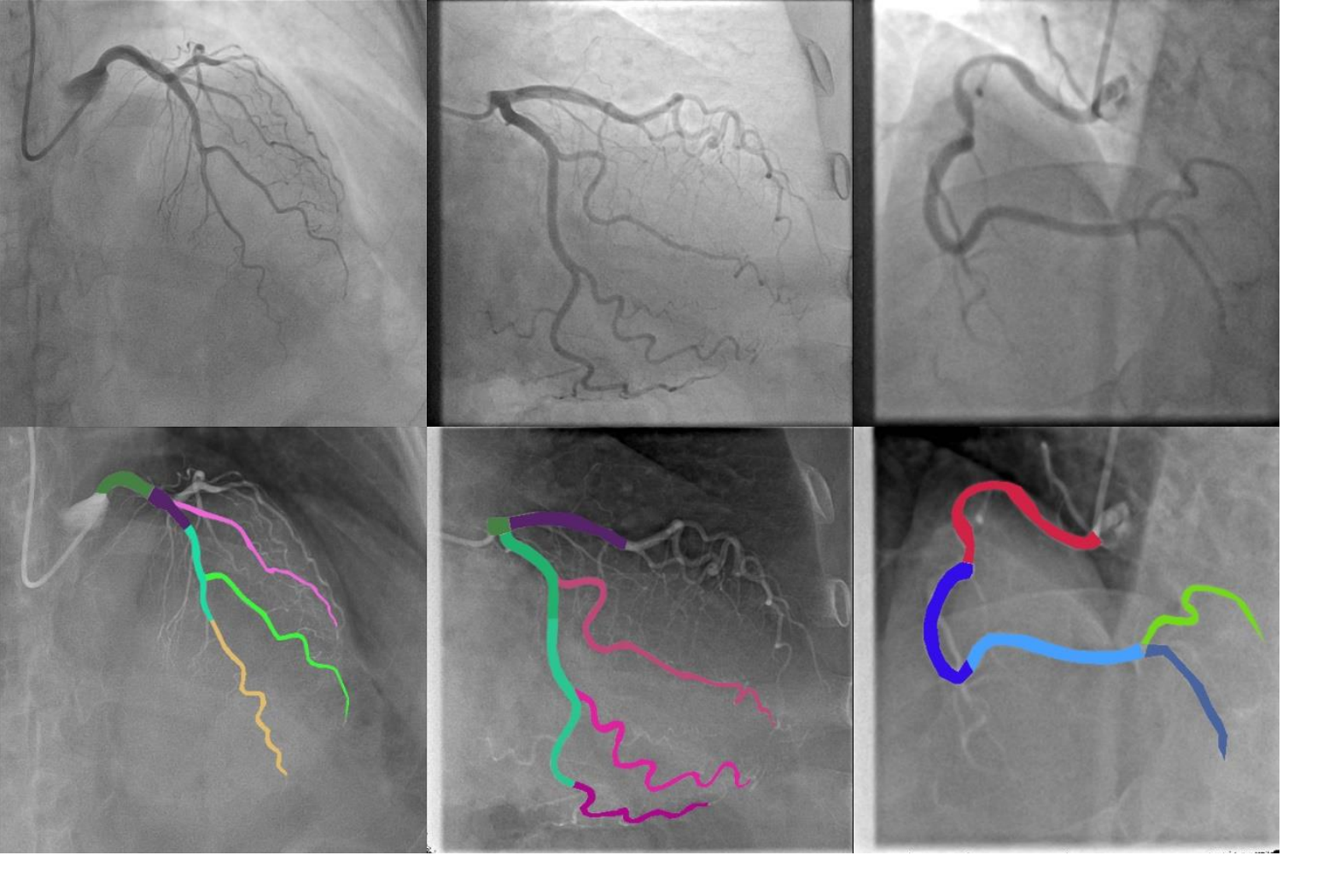}
    \caption{Segmentation of vessel into specific regions according to the Syntax Score. Top row represents sample angiographic images. Bottom row represents their division into regions according to Syntax Score guidelines. To the best of our knowledge, there is no publicly available model for multi class segmentation of vessel regions.}
    \label{fig:SYNTAX}
\end{figure}

Semantic segmentation tasks are aimed to perform pixel classification for a given image. In order to accomplish this task, several modern segmentation architectures have been proposed. The two most successful representatives are U-Net~\cite{ronneberger2015unet} and DeepLab~\cite{chen2017deeplab}. A large number of adaptations and modifications have also been proposed~\cite{deeplabv3plus2018,liu2019autodeeplab, wang2020maxdeeplab, qiao2021vipdeeplab, cheng2020panopticdeeplab, wang2020axialdeeplab, yu2022cmtdeeplab, meng2022unet3plus, Huang2020unet3plus,zhou2018unet++,zhang2018resunet,cciccek20163dunet,isensee2018nnunet,oktay2018attentionunet}. These approaches have succeeded for a large number of related tasks in medical image analysis, such as retinal vessel segmentation~\cite{li2019mresunetretinal}, renal ultrasound image segmentation~\cite{zhang2022deeplabkidney}, chest airway segmentation~\cite{garcia2019unetchest}, brain tumor segmentation~\cite{grovik2021deeplabbrain},breast tumor segmentation~\cite{negi2020unetbrest}, and many others. Inspired by their success, a series of meaningful improvements based on autoencoders have recently emerged, which improve segmentation quality. In the following, we will review a cross-section of these methodologies.

To date, only few papers compared the performance of various architectures for coronary vessel segmentation in a comprehensive manner~\cite{yang2019deep,Zhanchao2020MainCV}. Yang et al. compares the efficiency of different baseline models such as U-Net~\cite{ronneberger2015unet}, ResNet~\cite{he2016resnet}, Dense-Net~\cite{huang2017}, and InceptionResNet-V2~\cite{szegedy2016inceptionv4}. Their results indicate that DenseNet121 shows favourable performance coupled with a comparatively simple model, e.g. DenseNet121 has 5 times less parameters as InceptionResNet-V2, while exhibiting on-par performace in terms of F1 score. DenseNet121 reached 0.917 for this metric, while InceptionResNet-v2 scored insignificantly less – 0.915. The number of images used for training and testing all architecture is 3302 images from 2042 patients.

The network architecture comparison conducted in~\cite{Zhanchao2020MainCV} was done using a dataset of 3200 images, which were selected and annotated by cardiologists with an average experience of 10 years. Their study shown strong results for the Residual Attention Network~\cite{wang2017ResAttNet} for most metrics. Its average precision is reported as $0.919$ across different modalities, and DenseNet~\cite{huang2017} takes second place with an average precision of $0.912$. A disadvantage of this study lies in the fact that model complexity is not addressed.

A recent of application of conditional generative adversarial networks (cGANs)~\cite{du2021training} utilizes XCA images to automatically segment artery regions into 20 regions for pix2pix segmentation~\cite{du2021training}. The proposed cGAN consists of 2 networks – a generator and a discriminator. The generator is trained to generate an image from an input coronary angiogram, with the aim to reproduce the features of the original image, so that the discriminator network will classify whether each pixel is correctly labelled. The discriminator is, on the other side, is trained to distinguish between original images and ones produced by the generator. These two networks are trained until they reach a Nash equilibrium, where the discriminator can distinguish fake images from real ones with a probability of $50$\% (i.e. random guessing). Their work utilizes the biggest dataset used in this field of research - $20,612$ angiograms in total, with a segment recognition accuracy of $0.984$ and a sensitivity of $0.852$. An outstanding feature of this paper lies in its unique approach, which in some ways resembles the SYNTAX Score~\cite{neumann2018syntax}, inheriting the division of the coronary tree into specific regions, unlike most other segmentation architectures which employ a binary classification approach that merely distinguishes vessels and background noise. 

\begin{figure}[ht]
    \centering
    \includegraphics[width=\textwidth/2]{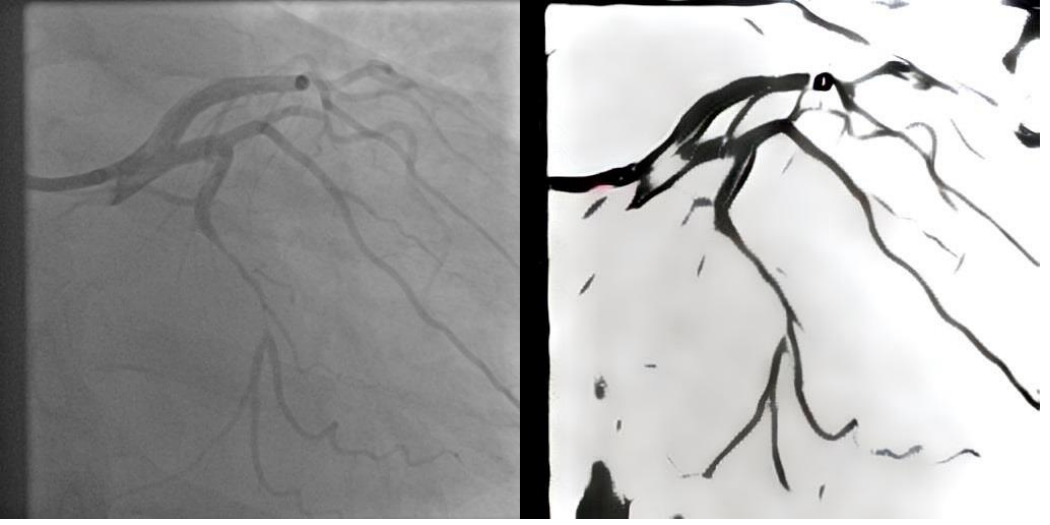}
    \caption{Sample segmentation using style transfer from open source black and white drawing using baseline AdaIN network. Original image on the left; Inference result on the right.}
    \label{fig:adain}
\end{figure}

Another creative approach to image segmentation using GANs was proposed in~\cite{Mulay2021adain}. The authors used a popular style transfer network with an Adaptive Instance Normalization layer~\cite{huang2017adain_orig} and added Convolution Block Attention Modules~\cite{woo2018cbam} and DexiNet~\cite{soria2020dexined} for improved style transfer performance. Uniqueness of this method lies in the extremely fast inference time and high accuracy of $0.9658$, which was tested using both open datasets described earlier~\cite{sanchez2019ann, hao2020svsnet}. This method does not require ground truth images, which were labelled by medical professionals for training purposes. Furthermore, it produces segmented images using style transfer from black and white drawings, which are taken from the BSDS500 dataset~\cite{arbelaez2011bsds500}. The result of style transfer using AdaIN network for XCA segmentation task can be observed on Fig.~\ref{fig:adain}.

A number of approaches have been proposed to incorporate the temporal dynamics of consecutive angiographic frames for segmentation~\cite{LIANG2021102894,hao2020svsnet}). Their idea was to explicitly model the flow of the color agent and thereby alleviate the negative impacts of artifacts and the patient’s movement on the segmentation performance.

Semi-3D TENet, proposed by~\cite{LIANG2021102894}, uses N=1 or N=2 additional images, preceding and following the manually selected best frame, resulting in a total of 3 or 5 images in the third input dimension. Angiographic videos, that were chosen for the dataset, can be divided into 2 classes – short-period videos, where the contrast-filled temporal area covers less than 10 frames, and long-period videos, where the contrast-filled temporal area covers more than 10 frames. To train and test this network $4904$ images were collected and manually annotated by medical students. $3099$ images were selected from the short-period videos, and $1805$ images were selected from the long-period videos. 
Comparative analysis of model performances after training on short-period and long-period videos separately has shown that the model achieves better score on short videos with N=1, while with longer videos better accuracy is reached when N=2. However, as there is much higher number of short-period videos in the training dataset, it was decided to use N=1 when training on the whole dataset (i.e. both short-period and long-period videos together). Results on the test set have shown the accuracy of $0.986$ and Intersection over Union score of $0.7221$.

The Deep Neural Network for sequential vessel segmentation (SVSNet)~\cite{hao2020svsnet} also uses the third dimension to effectively extract temporal correlation from the angiographic video. To verify the effectiveness of using the consecutive frame series as input, different numbers of input frames were tested (e.g. 2, 3, 4, 5, etc.). Experimental results proved that using 4 successive frames as an input resulted in the smaller loss and better model convergence. It was trained and tested using a total of 120 sequences of clinical XCA images. From these 120 sequences, 323 sample images were selected and manually annotated. The sample images include very low-contrast vessels as well as vessel trees that contain thin vessel branches, thus requiring the work of 3 experts to establish the ground truth. A more complex network, consisting of 3D convolution layers and Channel Attention Blocks (CAB) is reported to significantly improve performance as compared to naive 2D and 3D networks, with recall of $0.8424$ and F1 score of $0.8428$.

A novel ensemble framework, based on gradient-boosting decision trees \cite{friedman2001gbdt} and Deep Forest \cite{zhou2017deepforest} classifiers, developed in~\cite{gao2022vessel}, is reported to outperform U-net-based algorithms and modern DeepLabV3+~\cite{deeplabv3plus2018}. 21 filter-based features and 16 Dense-Net~\cite{huang2017} extracted features are combined into a 37-dimensional feature vector for each pixel of the XCA image. This ensemble method reached a mean precision of $0.857$, a specificity of $0.992$ and an F1 score of $0.874$ on a custom dataset of 130 manually annotated XCA images, of which 60\% were used for training, 20\% for validation, and 20\% for testing.

The idea of combining two different tasks to achieve better segmentation results is introduced and developed in~\cite{han2021jlnet,HAN2022106787}. By combining centerline and direction extraction tasks they achieve better vessel continuity for vessel segmentation. This resulted in increased sensitivity ($0.8561$) and F1 score ($0.8561$), compared to U-net and other modern segmentation methods with significantly lower number of trainable parameters (2.71 million) and only 200 images used for training.

\section{Lesion Detection, Localization, and Classification}

Object detection and classification are well-studied and established fields within computer vision~\cite{redmon2016yolo, wang2022yolov7, fu2017dssd, cui2018mdssd,shi2019ffedssd, sathish2020vggdssd, huang2018yololite,shafiee2017fastyolo}. For coronary angiography these tasks are currently performed in a semi-automated manner using a Quantitative Coronary Angiography (QCA) tool that comes along with medical imaging hardware. The drawbacks of this semi-automated process are that it is time-consuming and requires the cardiologist to manually highlight vessel structures. To date, a number of attempts have already been made to automate this process. Delegating the responsibility of stenosis labelling on XCA images to automatic algorithms would: i) serve as valuable assistive technology for the medical personel; ii) lessen the impact of subjective human perception and iii) increase the number of patients receiving timely treatment.

\begin{figure}[ht]
    \centering
    \includegraphics[width=\textwidth/2]{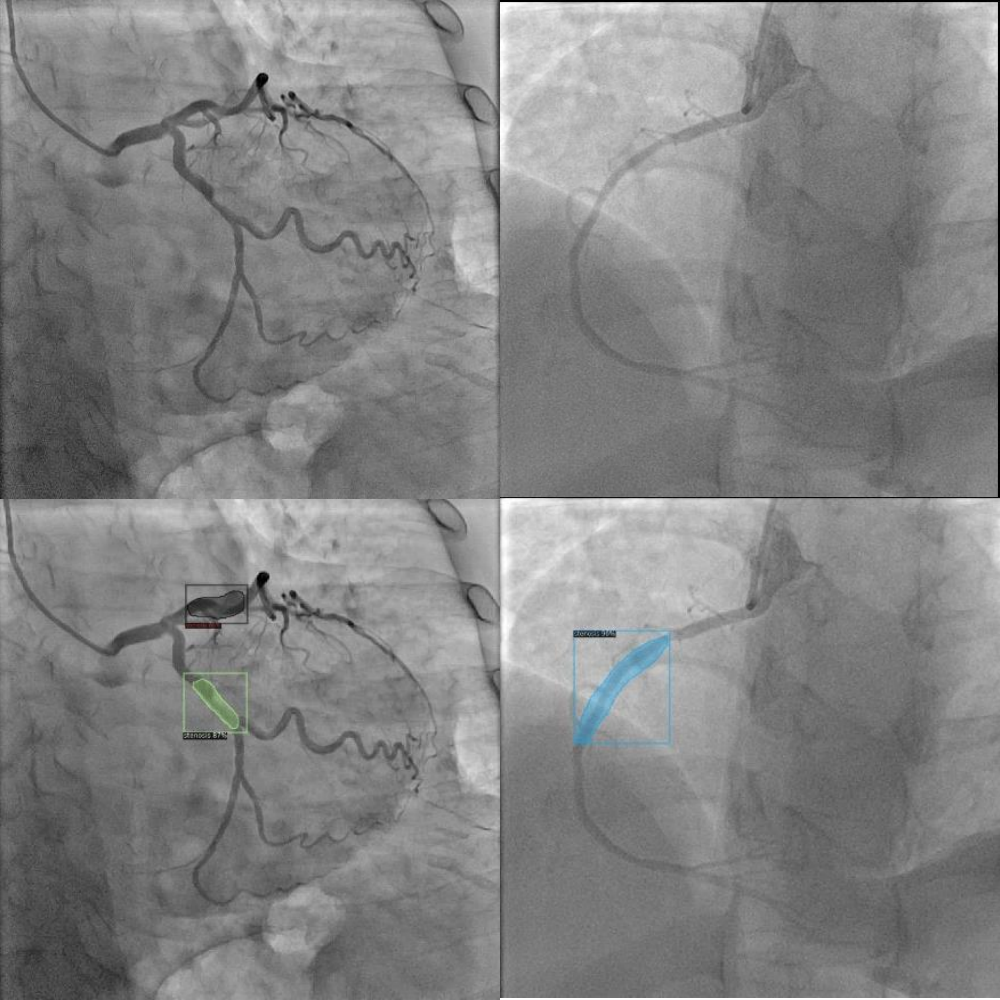}
    \caption{Detection results performed on the test dataset by sample network. To the best of our knowledge, there is no model trained and tested for lesion detection on coronary vessels available for public testing and benchmarking. Top row represents the original images, and bottom row represents model response.}
    \label{fig:detector}
\end{figure}

The Deconvolutional Single-Shot multibox Detector (DSSD)~\cite{fu2017dssd} was proposed for stenosis detection~\cite{WU2020103657}. This framework uses 3 adjustable hyperparameters, which are found during cross-validation. $N_cf$ is the number of contrast-filled frames, $N_{sf}$ the number of frames where stenosis occurs, and $N_{seq-iou}$ is the threshold for the Intersection over Union (IoU) operation for detecting whether detected stenosis occurs across several bounding boxes. The total number of sequences used in this paper is $148$, with 20\% of the samples used for testing. In this framework, the DSSD network makes a series of independent detections over several consecutive frames. The results of detection on independent frames are then compared, and if the result of detection is not persistent on all of the frames in the sequence, it is discarded, in order to reduce the number of false positive activations. These comparisons are carried out by a specific module called seq-fps. This approach proved to be a very effective solution, as without the seq-fps module, simple single-shot detection had a sensitivity of $0.871$ and Positive Prediction Value (PPV) of $0.461$, and after the processing with seq-fps the sensitivity remained unchanged, while PPV increased to $0.795$.

Pang~\emph{et al.} proposed Stenosis-DetNet for stenosis detection using bounding boxes~\cite{pang2021stenosis}. Their dataset consists of 166 images, of which 26 were chosen for testing, and the rest for training. In a similar way to Wu~\emph{et al.}~\cite{WU2020103657}, they propose a Sequence Feature Fusion (SFF) module, that allows to reduce the number of false positive responses of the network utilizing several consecutive frames for the detection, and one additional module, called Sequence Consistency Alignment module, that allows to confirm the location and class of a detected object using several consecutive frames. The combination of these two modules resulted in an increase of the precision to $0.9487$ and an F1 score of $0.881$, while method proposed by Wu and their colleagues got only the precision of $0.795$ on DetNet dataset.

A different approach to lesion detection is provided by~\cite{cong2019frame}. Instead of a detector network, classification network is employed, which is trained to classify images with $> 25\%$ of stenosis, and extracts an activation map from it.  The dataset contains 690 XCA images and was annotated independently by two expert cardiologists, with stenosis given a nominal ground truth bounding box of $35\times35$ pixels. Experiments were conducted in 2 different modes, with 3-class labelling, namely “$<25\%$, $25-99\%$, $>100\%$ Stenosis”, and 2-class labelling, namely “$< 25\%$” and “$> 25\%$ Stenosis” classes. The latter showed better results of the two in all metrics, reaching an F1 score of $0.76-0.83$.

The DeepDiscern network was proposed by~\cite{du2021training} and provides an automated tool for lesion detection and their localization. Du~\emph{et al.} trained their model with $6239$ samples and reached an average F1 score of $0.8236$ on the test set. One of the interesting and unique features of this work is that it discriminates lesions into 5 distinct classes: i) stenotic lesions; ii) total occlusion; iii) calcification; iv) thrombosis and v) dissection.

However, to the best of our knowledge, none of these fully-automated and powerful methodologies mentioned above are currently being implemented in clinical practice. Prior to their widespread adoption, large-scale open datasets need to be established and all methodologies need to be evaluated consistently in order to enable comparable and robust estimates of their quality. These types of benchmark datasets (e.g. MNIST, CIFAR, ImageNet, among many others) have shown to greatly accelerate scientific progress in many closely-related areas of computer vision as well as for machine learning in general, since they increase transparency and as a result the trust of scientific community. As a desirable future outcome, we hope that the creation of large-scale benchmark datasets for coronary angiography will not only be beneficial from a scientific perspective, but also increase the acceptance of clinicians towards these types of modern, fully-automated approaches.

\section{Open Problems and Limitations in the field}
As can be deduced from the analysis of the current state of the art in automated analysis of coronary angiograms, there are still several open problems that need to be addressed:
\begin{itemize}
    \item There is a shortage of large-scale, annotated coronary angiographic data, which have been labelled by experts of the field. While the above mentioned papers claim highly accurate results, their models are trained on either very small datasets  or they are trained on private datasets. This raises multiple concerns: i) the question remains whether these models are able  to generalize well to other data, which may contain diverse clinical cases of angiograms not included in the datasets availavle to the authors; ii) since most of the data is not publicly available, it is impossible to objectively verify the reported results and iii) it hinders many research groups to utilize this data and create their own models, thus limiting the number of approaches that can be employed. Producing large-scale data of high quality is difficult, since expert annotation of a given dataset for deep learning purposes is a time-consuming task and needs to be performed by professional clinicians, who have other clinical responsibilities and thus may be unable to annotate thousands of images in a short time.
    \item As for now, there is no real benchmark for angiogram abnormalities detection. Since datasets are either very small or private and most of the models (including their code) are not publicly available, standardized performance benchmarks are impossible. Similarly, there are no test sets for which proposed models can be evaluated independently. As a result efficient progress is hindered and there is a lack of consensus for building accurate and efficient models for this task.
    \item Some of the metrics, commonly used for ML-based model evaluation, are not suitable for clinical contexts. These metrics include sensitivity, specificity, accuracy and others related metrics, which are traditionally employed for the evaluation of ML models in general and DNNs in particular. However, these metrics deviate from ones that are used by medical professionals, who need to estimate the confidence of their diagnosis. This may hinder their acceptance by medical professionals and it is thus desirable to adapt the model evaluation strategies and their outputs to the professional's needs. Additionally, ML-based metrics may be complex to comprehend and interpret for some clinical specialists.
\end{itemize}

\bibliographystyle{IEEEtran}
\bibliography{ms}



\end{document}